# Computational Models
# of Spreadsheet-Development
## Basis for Educational Approaches

*Karin Hodnigg [1), Markus Clermont [2), Roland T. Mittermeir [1)*

[1) *Institut für Informatik-Systeme*
*Universität Klagenfurt*
*AUSTRIA*
*Roland.Mittermeir@uni-klu.ac.at , Karin.Hodnigg@uni-klu.ac.at*

[2) *Software Quality Research Laboratory,*
*Department of Computer Science and Information Systems*
*University of Limerick*
*IRELAND*
*Markus.Clermont@ul.ie*

**ABSTRACT**

*Among the multiple causes of high error rates in spreadsheets, lack of proper training and of deep understanding of the computational model upon which spreadsheet computations rest might not be the least issue. The paper addresses this problem by presenting a didactical model focussing on cell interaction, thus exceeding the atomicity of cell computations.*

*The approach is motivated by an investigation how different spreadsheet systems handle certain computational issues implied from moving cells, copy-paste operations, or recursion.*

## 1 INTRODUCTION

This paper departs from the perspective that spreadsheets are end-user programs. The main objective of spreadsheet development is "manipulation and presentation of data found in tabular form" [Filby, 1998]. The intuitiveness of spreadsheet development hides to a large degree that it is actually a programming activity. Typing constant values into some cells and a formula into another cell is not seen as programming. It is rather comparable to using a pocket calculator. The immediate presentation of the result even supports this notion. This allows introducing novices without much ado. One learns to use an environment instead of learning a model. While this can be seen as base for the high popularity of spreadsheet systems, it hides the reality that spreadsheet developers are expressing themselves in an inherently functional formula language.

[Nardi, Miller, 1990] identified immediate feedback through formula evaluation, tabular grid and related layout definition, the possibility to shift complexity by splitting formulas over different cells, and the rather declarative nature of most aspects of spreadsheet languages as sources of success. But while these features are certainly helpful for those writing small spreadsheets, they easily become obstacles when complexity increases. Certainly with large and evolving sheets, lack of higher level abstractions becomes a burden.

Only values are represented (formulas are eagerly evaluated), and formulas, but for a selected single cell, are hidden. This leads to a specific characteristic of spreadsheet pro-grams: hiding control and data flow information behind "static" values. During maintenance, this complicates comprehension of existing spreadsheets. Intertwining the layout of results and dependencies of computations is another source of conceptual complexity. Cells can reference each other over large geometrical distance. Hence, comprehension of a spreadsheet is a non-trivial task and consequently many errors are introduced or remain unnoticed. [Sajaniemi, 1998].

High error rates found in business spreadsheets ([Panko, 1998], [Mittermeir et al, 2002], [Brown Gould, 87]) indicate that the spreadsheet quality issue cannot only be resolved by powerful tools. One has rather to agree with Hoare's statement that "a significant challenge for programming theory is to […] develop an understanding to assist in the selection of an appropriate tool for each purpose." [Hoare, 1999].

Based on these reflections, this paper first identifies some crucial aspects of the spreadsheet paradigm. Then, it shows how differently basic issues are solved by different implementations of spreadsheet. This calls for a common conceptual background, which will be developed in section 5.

## 2 THE SPREADSHEET PARADIGM

To identify the target of this research, one is tempted to ask, whether there is such a thing as a "spreadsheet language" and if so, what this language might be. For a given product, it makes no difference, whether one types =*IF (A1 = B1; …)* or =*WENN( A1 = B1; …)*. These commands have the same effect on the data. Likewise, it makes no difference, whether this command has been typed in, selected by mouse click from some panel, or copied from a cell holding a similar formula which was edited afterwards. The clue is that the system provides the concept of an alternative and this concept is presented in different linguistic forms to the user. But the differences in linguistic form are rather shallow and users have to develop a conceptual model resting on the concepts behind the functions implemented in various spreadsheet products. Since all of these functions rest on common mathematical concepts, users must not be blamed if they assume that the sheet behaves in exactly the way they expect these mathematical functions to behave.

As the mathematics of the functions used in spreadsheets are well known to domain experts, the functional nature of cell-computations makes spreadsheet programming impressively simple [Moström, 1998]. Moström and Carr subsume the basic knowledge needed to implement a spreadsheet as follows:
- There is a tabular grid consisting of (addressable) cells.
- A cell can hold either a formula or a static value.
- The formula language is declarative, having the form
    =*<cell_addr> [ <operator> <cell_addr>]* or
    =*<function>(<cell_addr1>,..,< cell_addrn> [;<system parameter>])*
  with system parameter being an element of the spreadsheet system rather than the mathematical base of the language.
- Cells can be referenced as solitaires (*A1*) or by range reference (*A1:A12*). A reference can be either absolute or relative.

This resembles functional programming and clearly contrasts with "conventional" programming. The spreadsheet language is mathematically traceable and highly declarative as it "emphasises on the evaluation of expressions" [Montigel, 2002]. Normally, spreadsheet languages focus on spatial relations of data, not on the temporal sequence.

This applies at least to very basic concepts for implementing spreadsheet programs and lead to the statement that the spreadsheet language is a "programming language for the masses" [Moström, 1998]. Even without specific training, everybody can write models based on the writers domain expertise. But there is no warning when those limits are left because extensions of the spreadsheet paradigm like those discussed in section 4 violate basic assumptions of the model.

**The Conceptual Mismatch**

For a novice, developing a spreadsheet is like a child's building a play-house. Formulas are written into cells like placing bricks on bricks. The intermediate results can be admired after each step. With sufficiently small problems this cell-by-cell approach can follow an implicit DDG almost in breath-first manner. The geometrical placement of cells is almost irrelevant.

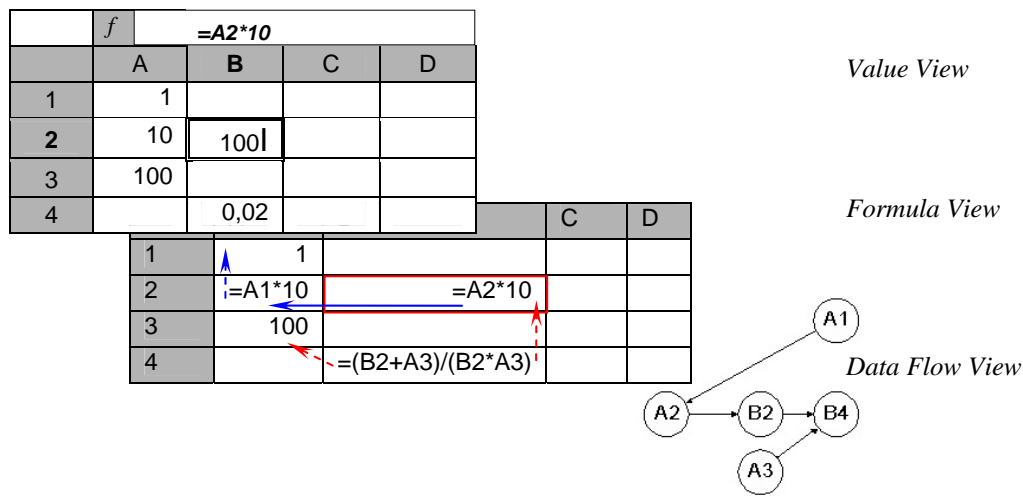

**Figure 2: A user's view based upon [Igarashi, 1998]**

But what is easy during construction might prove to be difficult for later comprehension. When perusing cell *B2* in Fig. 2, the cell's value is shown in the sheet and the underlying formula is displayed in the formula bar. The constant values in *A1* and *A3* and the formulas in *A2* and *B2* are related by invisible dependencies. Clicking into one of these cells, the spreadsheet GUI shows the "first level"-dependencies, i.e., the addresses of source cells. A spreadsheet programmer may notice that *A2* depends on *A1,* but which cells depend on *B2* remains totally hidden. Thus, changes to *B2* may result in changes "anywhere". Theoretically, any computation might be affected by changes in a given cell.

Navarro-Prieto and Canas' results indicate that spreadsheet writers have developed good mental structures for data flow information [Navarro-Prieto, Canas, 1999]. [Tukiainen, 2001], in contrast, points out the need to memorize (invisible) coherence in a spreadsheet without explicit representation. We conjecture that comprehension becomes increasingly difficult with growing size of the sheet. It will matter specifically, when references exceed the window visible on the screen. Assuming maintenance of the sheet, the situation is aggravated. With fading memory, the basis for Navarro-Prieto's hypothesis is lost.

Another issue exhibited by Fig. 2 is a likely conceptual mismatch between the displayed values and the underlying calculation model. The user does not "see" the data dependencies. They are hidden behind (possibly misconceived static) values. For a spreadsheet maintainer a visit of every content-bearing cell is necessary to build a mental model of the spreadsheet program. Every visit extends the user's conceptual model, thus only the "final" (i.e. the most recent) model corresponds to the real data flow of the program. The fact that users see data dependence at best incrementally is particularly detrimental when those are not in line with the generally assumed left-to-right, top-to-bottom assumption and when the value-perspective allows several data-flow interpretations.

## 3 SPREADSHEET PROGRAMMING CHARACTERISTICS

Though intuitive, assumed data flow semantics reach their limits in critical situations. As stated in [Clermont, 2003], spreadsheet programs "share many features with data flow concepts", but "some of the key concepts, such as the consumption of tokens" are not part of the spreadsheet paradigm. Questioning this statement, one realizes that spreadsheet semantics are tool-dependent. Established models to describe programming language semantics are not fully applicable and hence the broadly shared interpretation of spreadsheets as dataflow programs breaks down.

### 3.1 Evaluation Strategies

A decisive difference between conventional (functional) programs and spreadsheet programs is evaluation time and process. Conventional programs are fully specified before they are evaluated whereas spreadsheet evaluation takes place after each incremental development step. The end of the development process is never made explicit to the system, though. It stops, when the developer is satisfied with the results of the computation.

So far, we have been unspecific when referring indiscriminately to the functional paradigm and to the data flow paradigm. Addressing evaluation, though, one has to recognize that these concepts differ in evaluation order and concept.

**Data Flow Semantics**

Data flow programs (DFP) like spreadsheet programs do not need an explicitly defined control flow. Order of execution is implicitly defined. In DFP, data dependencies control the sequence of function evaluation. As a data flow program is usually visualized by a data flow graph, one may conceptualize evaluation of a given node as soon as all its edges bear data, i.e., all required information is available. The data propagated is a token that holds the result yielded by the computation of the node placed immediately downstream. This evaluation concept is based upon the consumption of tokens. Re-evaluation of the DFP implies re-computation of all tokens.

With spreadsheet semantics, however, there is no "value" marking of these edges of the DDG but rather a marking of "change". [Yoder Cohn, 2002] point to this crucial difference: In a data flow program, a cell is re-evaluated only, if *all* of its sources have new values for processing. In spreadsheets though, a single re-evaluation marker suffices to trigger re-evaluation. This marker is an explicit element of control: Moreover, for treating loops, DFPs include loop nodes as special concept. But loops are not part of the standard spreadsheet paradigm.[1]

---

[1] ) Excel allows recursion with limited iterations (see 4.2). But this is rather not part of the standard repertory of spreadsheet writers and has conceptual limits which may result in unwanted side-effects.

**Graph Reduction Semantic**

Passing control seems akin to functional programming with its graph reduction semantic [Sestoft, 2001], [Dermoudy, 2003]. Here, each formula is interpreted as functional statement. Graph reduction semantics imply that a function call and its arguments are replaced with the result of function application. Since the result of a function can be used more than once in a program, reduction has to be repeated for each occurrence.

Consequently, [Clermont 2003] postulates that spreadsheet program evaluation seemingly follows graph reduction principles. There are two main arguments that show that spreadsheet programs are no pure graph reduction programs though: loops and change propagation.
- *Recursion and Loops:* The functional programming paradigm does not include loops. Recursion is the concept to express repetition. Recursion, however, is not part of the spreadsheet paradigm[2] since it inhibits the visibility of intermediate results and postulates inherently the provision of a global control flow.
- *Change propagation:* The interactivity of spreadsheet programs leads to a sophisticated change propagation technique [Clermont, 2003], [Yoder Cohn, 2002]. If a cell's content changes three steps happen to maintain consistency:
  - the formula's value has to be re-evaluated,
  - depending formulas have to be re-evaluated, and
  - formulas within the transitive closure have to be re-evaluated.

Thus, re-evaluation is mainly token-driven. [Burnett et. al., 2001] coined the term "continuous evaluation" to highlight immediate currency of results.

**3.2 Inconsistent Evaluation Strategies**

To resolve these contradictions, [Clermont, 2003] suggests spreadsheet programs to be considered partly as graph reduction program and partly as data flow program. Which of the two applies in a given situation becomes relevant during spreadsheet maintenance. A distinction has to be made between local and global evaluation:
- *Local Evaluation:* Evaluation of a cell's initial value (starting from its formula) is based upon graph reduction according to the spreadsheet's DDG.
- *Global Evaluation:* If changes in a cell occur, they are propagated via the data flow graph to all dependent cells through change tokens. This data flow graph corresponds to the reverse DDG of the spreadsheet. Thus, cells are only re-evaluated when needed.

Hence, spreadsheet programs incorporate both concepts, depending on viewpoint. Due to interactivity and visibility of all cells, the global evaluation strategy implemented has to be eager. However, this does not prevent lazy evaluation to be locally applied, e.g. Excel evaluates IF-clauses lazily.

There is another distinction between functional and spreadsheet programs. By the nature of the spreadsheet paradigm, every cell on the spreadsheet has to be considered as output whereas a functional program has a set of selected outputs. In [Yoder Cohn, 1994] an approach is presented that is based upon demand-driven evaluation of the cells of interest only. But the authors rely on "keeping all cell values up to date".

Looking behind these differences, a spreadsheet program unifies functional and data flow concepts. As stated in the references, the very (natural) base of functional languages is data flow graphs where functions are nodes and edges represent the data dependencies.

---
[2]) For exceptions and approximations see section 4.2.

## 4 IMPLEMENTATION DIFFERENCES

While section 3 discussed principles of spreadsheet evaluation this one concentrates on actual implementations of spreadsheet systems. How do they treat evaluation and where are differences between systems or between concept and implementation? Apparently, such differences will constitute risks for development and pitfalls for education.

### 4.1 System Specific Evaluation Strategies

To analyze how system builders resolve the crucial design issues for an evaluation strategy, reducing unnecessary re-computation and maximizing use of available computational resources [Yoder Cohn, 1994], the implementation of Microsoft's Excel$^©$ and Gnumeric are used. For the open source product Gnumeric, the sources used are quoted in the discussion. For Excel (2000) we had to rely mainly on the Online-Help.

**Excel Value Recalculation and Change Propagation**

To be efficient, Excel performs so-called "minimum recalculation" using the following strategy [La Penna, 2001]. It keeps an internal list of all linked (interdependent), cells bearing formulas in a workbook (like a2>b2>b4). Cells with constant values are not part of the list, as they cannot be affected by change propagation. If a change occurs, all cells (transitively) dependent on the cell that changes are marked with a recalculation flag. Recalculation starts according an internal list of dependent cells.

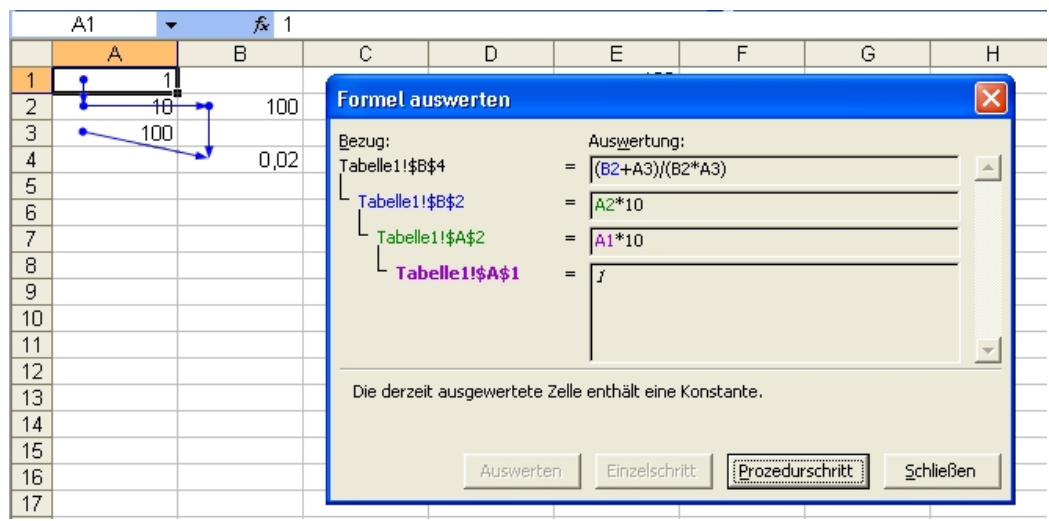

**Figure 3 Formula Evaluation in Microsoft Excel 2003**

Fig. 3 shows Excel 2003's Formula Evaluator applied to the sheet of Fig. 2. The evaluation path starts at the formula in *B4*, a data flow sink in this sheet. It continues in a stepwise manner against the direction of data flow to cells *B2, A2*, till *A1*, a constant. Thus, graph reduction is performed and every single step can be seen by the user.

**Gnumeric Value Recalculation and Change Propagation**

Gnumeric distinguishes between two types of dependencies: single dependency, which is a reference to an individual cell (*=A1*), and range dependency, which encompasses a set of cells (=*SUM(A1:A5)*. Since it is necessary to determine dependencies for every (transitively) dependent cell, both dependency types are mapped to distinctive data structures to ease the lookup of a cell's dependents.

To recalculate values, a recursive approach traverses the given expression tree, re-calculating each source cell (recursively, if the source cell contains a formula). This string reduction approach is rather inefficient if the same cells are to be re-evaluated more than once. Hence, Gnumeric buffers the cells in an evaluation queue, traversing the dependency data structure. Value recalculation then corresponds to graph reduction manner

### 4.2 Circular references and Iteration

The spreadsheet paradigm does not provide any kind of global control flow. Thus, loops (iterations) and recursion are not defined. There are spreadsheet systems that simply prohibit (even the accidental use) of circular references. Francoeur states for an *ExcelComp*-Tool that "an admissible spreadsheet contains no directed cycles", i.e. no recursion [Francoeur, 2002]. Accidental use happens though by incorporating the cell holding an aggregation function itself into the scope of aggregation e.g. writing *=SUM(A1:A3)* into *A3*. Surprisingly enough, there is no common approach to handle these circular references.

**Microsoft Excel Circular References**

Excel provides an ignorable warning. If a circular reference is accepted "as is" then every cell containing a function leading to the circle is considered to be a terminal node (constant). Neither this cell nor its dependents will ever be re-evaluated after the warning has been ignored. Without any special marking, the cell provides the value zero. This is quite problematic since *"0"* could be a legitimate value expected by the user and the rupture in the evaluation path will lead to wrong results anywhere. Dependent values will remain on whatever value they had before the recursive case occurred and remain so, even if any of their other sources is changed.

By a special command-panel Excel provides an iteration scheme exceeding the spreadsheet paradigm. It allows to evaluate a cell containing a circular reference over a user defined constant number of iterations. If a change anywhere in the sheet, even outside the transitive closure of the cell's sources occurs, the cell is re-evaluated. Excel offers this to accommodate requirements of some scientific computations. Nevertheless, the introduction of an even reduced iteration model is a substantial intervention into the "traditional" spreadsheet paradigm.

**Gnumeric Circular References**

Gnumeric includes circular references into its concept. No warning or indication is given to the user. Gnumeric does not consider circular referencing cells being terminal nodes. In some (!) cases it treats circular references as a kind of two-staged loop. If the value of a cell with a circular reference has to be evaluated, Gnumeric supposes it to start with a given value zero (0), computes the function over the non-recursive part and takes this value to recompute over the full extent. If recalculation of that cell is necessary, i.e. if changes in one of its source cells occur, Gnumeric consults the given (old) value and uses it for re-computing the new result.

A small experiment shows this behaviour: *A1* holds a constant value (*1*); *A2* builds the sum of *A1* and itself *(=A1+A2)*. After the input of this formula, Gnumeric yields *2* in *A2*. If the value in *A1* is changed to *4*, *A2* becomes *10*. If subsequently *A1* is changed back to *1*, *A2* becomes *12*. Thus, the "previous" value of both cells and the "new" value of the changed cell are used to build the sum: (*2 = 1+1+0, 10 = 4+4+2, 12 = 1+1+10*). I.e., a non-recursive computation is performed and its result is added to the previous value contained in the recursive cell. Aggregation functions such as *AVG*() and *SUM*() provide results in a similar way. Trying to incorporate subtraction in *A2* *(=A1-A2)* leads, unexpectedly, to *0* with no re-evaluation taking place though. With this feature, the treatment of circular references seems even more dangerous, since probably incorrect values are computed without any warning.

**4.3 Copy/Paste and movement heuristics**

A discussion of spreadsheets will be incomplete if specialties of the development process such as "drag-and-drop" or "copy/paste" are not considered. As formulas are parameterised by (either constants and/or) relative cell references, the way these references are adjusted is crucial for spreadsheet correctness.

**Moving Cells**

Moving cells from by a drag-and-drop operation is a common operation in spreadsheet development. It is distinct from cut and re-paste at a different location, since by drag-and-drop the link to the referenced cells persist while cut-and-paste preserves the geometrical pattern of the relative addresses. To keep the spreadsheet consistent can be resolved in two ways. References pointing to the "moving" cells could move with the cells (according to a pointer idea) or references starting from the cells could rather keep the reference treating them as a "geometrical" pattern. In the latter case, the DDG will remain unaffected except that the moving node will get a different address-label.

Both systems, Gnumeric and Excel provide this feature with computational reference keeping dominance over geometrical patterns. This principle is implemented for both, moving cells, and inserting columns or rows. The deletion of cell block contents leads to zeros in cells that refer to the removed block, since the cells exist but do no longer hold any value and zero is considered as default value for empty cells. If whole columns or rows are totally deleted, though, a reference problem pops up as the referenced cell does not exist any more. In this case Excel displays the error value *#REF*, Gnumeric does not display a value, although the cell carries the *#REF!*-value. Aggregation functions such as *SUM* or *AVG* play a special role in this context though. The range covered by these formulas dominates over reference or geometrical patterns.

Figure 4 highlights the interaction between development steps and aggregation functions. Cells *A1* and *A2* contain constants. *A3* and *A4* contain the formulas *=A1\*10* and *=A2\*10* respectively. *A5* contains an aggregation function with the range *A1:A2*. If the formula-block *A3:A5* is copied to column *C*, the references are adapted according to the geometrical pattern of their copy source. If the same block is moved from *A3:A5* to column B though, the references to the source cells persist. If a referenced cell, say *A2*, is moved to *D2*, then the referencing cell (still *A4*) adapts and in a notion of pointer semantics points to the new cell containing the referenced value. Thus *A4* contains the formula *= D2\*10*. Interestingly enough, the aggregation formula does not adapt as well. *A5* still contains the formula *=SUM(A1:A2)* and yields the value *1*. The content of cell *A6* will in this case yield *31*. Apparently, history is only partly preserved and the notion of physical areas dominates over development history.

|   | A | B | C | D |
|---|---|---|---|---|
| 1 | 1 |   |   |   |
| 2 | 2 | =A1*10 |   |   |
| 3 | =A1*10 | =A2*10 | =C1*10 |   |
| 4 | =A2*10 | =SUM(A1:A2) | =C2*10 |   |
| 5 | =SUM(A1:A2) |   | =SUM(C1:C2) |   |
| 6 | =SUM(A1:A4) |   |   |   |

**Fig. 4: Copy and Paste vs. Drag and Drop**

### 4.4 Filling Cells

All spreadsheet systems provide "filling" operations. Starting from a given cell users can automatically "fill" geometrically neighbouring cells either with values or formulas. In all cases, default adjustments are made. Whether these defaults are intuitive and meet the developer's expectation depends on the situation and on the developer's conceptual model.

**Filling Cells with Values**

Automated filling with values seems straightforward. Nevertheless, there are some differences between the systems considered. Both Excel and Gnumeric provide a "copy of constants" operation to duplicate the value of a single starting cell and a "series copy" operation that successively increments values (e.g. *a2=a1+1*). The distinction (common to every spreadsheet program) is based upon the "Control"-key.

If more than one value is selected as starting point to the value series, Excel tries to figure out the subsequent values by building a geometric series. So, if a user wants to fill a block with four values down a column, let them be (*3, 7, 2, 5*), a geometric series is built up (*3, 7, 2, 5, 4.5, 4.6, 4.7, 4.8, 4.9, 5, 5.1, 5.2, 5.3, 5.4, 5.5*). This may not be exactly what the user expects. A copy operation with the constant values (*3, 7, 2, 5, 3, 7, 2, 5, …*) is triggered via the control-key, which is in this case non-intuitional.

Gnumeric on the other hand takes the differences of the last two values of a block to compute sequent values. The filling of a column with the values (*3, 7, 2, 5*) leads to a sequence of following values (*3, 7, 2, 5, 8, 11, 14, 17, 20, …*). In Gnumeric, the control-key does not provide another function.

**Filling Cells with Formulas**

When clicking on cells filled with formulas and dragging them over an area, references are treated as geometrical pattern and constants maintain their value. Thus, there is no adaptation as discussed above. If more than one formula is selected to be copied, the block of formulas is taken and cloned, if the selection window is dragged over either rows or columns. Hence, different constants in two successive formulas remain different in the newly filled formula block. In [Igarashi et al., 1998] an interactive graphical induction approach is presented. There, the structure of a spreadsheet program and regular patterns are used to induce a continuative formula pattern.

### 4.5 Conclusion on differences

The comparison showed that even such typical spreadsheet operations like movement and filling of cells by "drag-and-drop" are implemented differently in frequently used sys-tems. This can cause unexpected results. Even if differences seem to be marginal, they indicate that common spreadsheet didactics is hard to achieve and system specifics have to creep in. Worst about these differences is the marginality of distinction. Different spreadsheet systems behave identical in most situations, but not always! One has to conjecture that there is no common spreadsheet language which can be defined as the union of a distinct formula language and tabular layout issues. To fully understand spreadsheet development, one has to learn details of the system too. Moreover, adaptation heuristics, defined to help spreadsheet developers, cause effects of non-associativity in the sequence of certain development steps.

## 5. SPREADSHEET SEMANTICS

The previous section has shown limits of comprehending spreadsheets and hence of teaching spreadsheet development on a cell-level basis. The naïve perception of spreadsheets as an arrangement of cells (c.f. the bricklayer's approach described in sect. 2) reaches its limits either when computations become too involved or when due to maintenance operations incorporating new requirements the sheet evolves over time. Here, a model is introduced that should help to comprehend the interdependencies between cells without falling into the problems mentioned for data flow semantics or reduction semantics discussed in section 3.

As any model describing the semantics of a language, a model for spreadsheet semantics has to be expressive and faithful with respect to the intended semantics spreadsheet users and developers of spreadsheet systems have in mind. Further, such a model has to be simple in so far as it requires only a minimal number of primitives. Finally, considering the spreadsheet user-community, it has to be highly intuitive. The latter argument is to be seen as a distinction between programming language semantics and spreadsheet language semantics, since the former are to be understood by programming professionals whereas the latter are to be understood by application experts who are rather programming laypersons.

As shown in section 4, the divergent semantics of spreadsheet systems conflict with a notion of common spreadsheet semantics. Hence, one must not expect that such a model covers all detailed variations implemented. However, even if it does not cover them, it should at least not be in conflict with them.

### 5.1 Relationships and Visibility

Another peculiarity of spreadsheets is to be considered. According to the bricklayer-semantics, each brick (cell) can be placed anywhere on the sheet. (Transitive) dependencies are established due to the relationship(s) a formula establishes with the cells it references. These references are normally represented as relative positional distances from the target cell to the source cells. Hence, whenever the target cell moves, the references to the source cells experience an identical movement irrespective of whether the respective positions in the sheet contain appropriate values or not. Absolute references, i.e., references to fixed positions are also possible but not the norm. However, even in these cases the absolutely referenced cell can be anywhere on the spreadsheet. Cells serving as parameters (or data sources) for a particular formula can be arbitrarily spread over the sheet. Modifying the spreadsheet program by inserting or deleting rows or columns affects neither relative nor (interestingly!) absolute references. The relationship once established to a particular cell (with its content, either constant value or computed value) remains. Thus distance (relative) or address (absolute) is adapted as shown in section 4.3.

The relationships holding in the computational perspective of the sheet are stronger than the positional aspects.

This freedom is lost with aggregation functions. For them, the concept of a (physical) area has been defined. This is a rectangular block of horizontally and vertically consecutive cells. In this case, however, the target cell receives its value from cells placed within the geometrical confinement of this rectangle. Deletion and/or insertion of rows or columns may affect this area, if they take place within the borders of the area (not, if they take place at the border). Thus the area has a certain degree of flexibility. Further, areas yielding results into aggregation functions placed in different cells might overlap (which would be a contradiction with pure data-flow semantics). Thus, there is, like with individual source cells, no unique ownership of sources.

While cells on a yet empty sheet and cells containing only constant literals are globally visible indeed, this does not hold for cells containing formulas. A cell holding a formula that references only constants might still be conceived as globally visible. However, this cell cannot "see" any cell that directly or indirectly serves as target for the value of its own computation. Otherwise, the computation would contain circular references.[3] Thus, there is an implicit visibility arrangement between cells. This arrangement depends on the (transitive) target-source relationship between cells.

**5.2 The Projection-Screen Model**

Computations confined to individual cells are not a problem in spreadsheet education, since they are conceived as functions well understood in the application domain. The functional nature of cell-based computation provides clear scoping. The global model remains conceptually unsupported though. It is not adequately addressed in introductory teachings, conventional models cannot fully account for all effects, and typical spreadsheet operations might conflict with them.

Therefore, we present a model based on the interrelationship of cells. Drawing upon the instant visibility of results of any computation and the implicit relationship between cells due to data dependency, we interpret cells as optical devices, reading results from screens (cells) placed in front of them and projecting the result of their computation on their own screen which supposedly is placed on the back of the viewing mechanism.

**Projection-Screens without aggregation**

Spreadsheets containing just empty cells, cells with constants and cells with non-aggregative formulas serve as point of departure. For all cells hold:
- ***Empty cells*** can be ignored, since they do not partake in any computation.
- Cells containing *constant literals* might contain labels or constants to be used in computations, i.e. by other formula cells.
    o ***Labels*** do not partake in any computation. Hence, one might be tempted to treat them like empty cells. However, not the cell holding a literal decides on its usage. A literal cell is globally visible. Hence, any other cell in the sheet can at any time in the development process reference this cell. Hence, labels are treated like computational literals.
    o ***Computational literals*** (usually numeric values) are treated as primitive 0-argument formulas. Their result is the value denoted by the very literal. *0-*

---

[3] ) For exceptions see section 4.2, treating circular references in different implementations.

*argument formulas* (e.g. *NOW()*) look at no other screen. They only present their own value on their own screen, thus making it globally visible.
- **Non-aggregative formula** *cells* have as many arguments as they are (relatively or absolutely) referencing cells in their formula. They read in a non-destructive manner the values from screens mentioned in the formula and project the result of their computation on their own screen. The screens they read from have to be conceptually positioned "in front of them".

This model requires a unique viewing direction among screens. But it is completely independent of geometrical placement of source or target cells and neutral with respect to computations referencing individual cells. It shares directionality with the data-flow model, but in contrast, nothing flows. Formula cells just read from the projection-screens "in front" of them and hide their own results from screens "in front." As the viewing devices are constantly attentive, they realize when a value of one of the screens in front of them changes. This leads to re-evaluation of the own formula. Thus, intermediate results, whether they stem from literals (0-argument formulas) or regular formulas, can be shared by as many target formulas as needed as long as the direction of visibility "look in front of you, write the results on your back" is upheld.

However, from a risk assessment perspective, one might check, whether the direction of visibility can be linearly mapped to a partial order in the geometric placement of cells. Deviations might serve as complexity measure. Computation of related risk indices would go beyond the scope of this paper though.

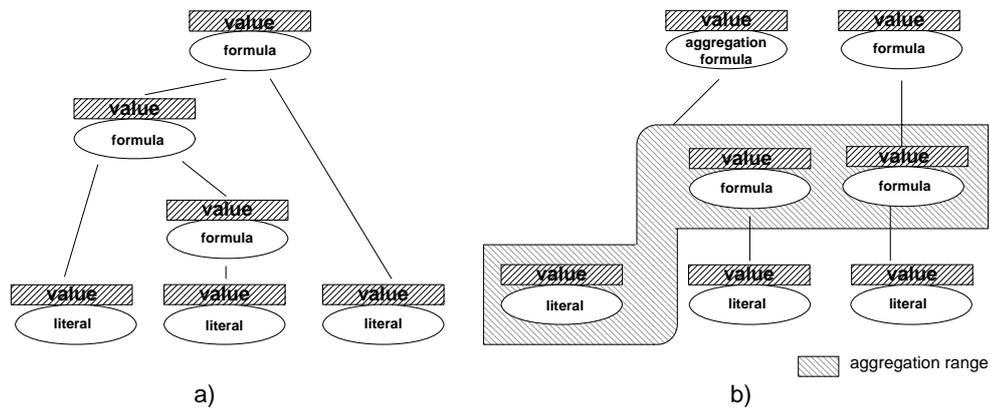

**Figure 4: Projection Screens a) without und b) with aggregation functions**

**Projection-Screens with aggregation**

One might be tempted to treat aggregation functions simply as shorthand for explicitly mentioning a (huge) set of arguments. But section 4.3 shows that aggregative and non-aggregative functions differ from an evolutionary perspective. Thus, the semantics of aggregation needs special treatment.

Still resting on visibility, one might consider the area under aggregation as the set of screens illuminated by a common spotlight. This concept withstands evolution, since deleting a portion will reduce the area of visibility, while inserting empty space will enlarge it. If, later on, this empty space is filled, it is seen by the aggregation-function's viewing mechanism as if it had ever been there. Like on stage, the aggregation function sets a spotlight on that ensemble of actors (still all looking up-front, carrying their result clearly displayed on the screen carried on their back) that partake in this particular aggregation. Like on stage, different spotlights (on the stage they have possibly different colour) might illuminate different actors (cells) and some of them might be in the focus of different spotlights. Hence, they yield their values for different aggregation functions. Actors leaving the illuminated areas are no longer seen by spectators.

### 5.3 Discussion of selected evolutionary steps

For the sake of demonstration, some prototypical patterns frequently recurring on spreadsheets are analyzed with respect to the projector-screen model.

**Many-handed figures**

Certain goddesses such as Bodhisattvas are depicted with many hands reaching out in different directions. Reducing this to spreadsheets amounts to a cell (or block of cells) affected by evolution that has several dependencies outside of the block manipulated. In its most concentrated form, one might conceive of an *IF-statement*, consisting of *<condition>*, *<argumentT>*, *<argumentF>* where each of the three formulas might have references to other cells that extend over the geometrical area affected by movements, insertions, or deletions. As long as these manipulations do not affect a cell directly addressed in this "many-handed-statement" (which could happen with deletions or with movements over an area where such a source-cell is located), the computations in the statement are not affected. The principle that computational references dominate over location-based references applies. Thus, not only relative addresses are adjusted properly. Even absolute references are adjusted to maintain established data connections.

This is consistent with the projection-screen model. Visibility is strictly based on position of the screen relative to the screen it reads from. On the first level, those exhibit only constant values (either input cells or cells used as constants of the spreadsheet program). Every formula computing intermediate results used by a cell on the path between the panel of constants and the cell considered puts this cell one level towards the rear of this screen/projector scene. But as insertions of rows or columns have no effect on this distance nor do deletions of rows or columns (that do not directly affect a node on this chain) have any effect on this arrangement, the projection-screen model is consistent with such operations. The same applies if not a single cell but a block of cells is considered. Even if insertions (deletions) modify the size of this block, no changes in screen positioning are needed and, therefore, no visibility changes are induced.

**Queue on a staircase**

The previous case considered independent cells or independent areas. There are cases with connections within a block of cells that is treated as conceptual entity. Running numbers are the simplest example. One starts with some constant, say *1* in e.g. cell *C3*, writes *=C3+1* into cell *C4*, and fills cells *C5* till *Cnn* by dragging down *C4*. Right to these running numbers, usually information with application semantics is given. It might be necessary to insert or delete a row or to move part of this construct to some other place. Although the block with running numbers is a conceptual entity, the spreadsheet system deals with it as set of neighbouring cells with each one having just one external reference. Hence, the rules for the many-handed figures (here: single-handed) apply.

This also holds for the projection-screen model. One might envision a staircase where the front element holds the constant (here *1*) and displays it on the screen on its back. All other elements are looking at the (single) screen immediately in front of them and display whatever they read incremented by *1*. Adding a step to this staircase, or moving the tail of the queue some steps back (or the front of the queue some steps forward) does not change this visibility. Geometrical reference is adjusted to maintain the relationship to the visible screen in front of the viewer directly affected by the geometric rupture. Deletion of individual cells, however, does have an effect in this case, since not only the step of the deleted is removed; its screen is also blinded. Hence, reference is lost and #REF is displayed as error message for the cell that lost its ancestor, but also for the cells "in the back" of the respective cell. However, when the head of this affected sub-queue gets fixed, its tail and thus the complete queue is fixed automatically. The general visibility system and the computation mechanism of the dependent cells is not affected in this case. Those cells just cannot produce interpretable results because (one of) their ancestor(s) shows no result on its otherwise perfect screen.[4]

But what, if not the cell but just the formula is deleted? In this case, the queue starts with *1* again without reporting any problem. Is this consistent? It is! Due to the default value *0* for empty cells, the cell behind the empty one notices *0* in the predecessor, increments it and thus displays *1* and all cells behind it act accordingly. Thus, a new queue is defined.

**Flying carpets**

Finally, one should look at aggregation functions. Here, the cells to be aggregated over are affected by operations relating to the geometrical arrangement of the sheet. But the cell containing the aggregation function can, like a flying carpet, be freely moved to a new geometric position without loosing sight. One might assume that this causes problems in a model relying strictly on data dependencies. The spotlight-interpretation of visibility helps though.

The spotlight covers an area (on stage as well as on the sheet). This illuminated area is independent of whether the area is populated or not. The viewing mechanism has always to be in the back such that all items in the illuminated part can be seen. Thus, conceptually, it might be necessary to step back, if something is inserted that is already at a level far away from the front panels showing constants. If maintenance operations change the size of the illuminated area, it is important to note that the scope of illumination is always defined by the fringe positions. This border does not change. Hence, deletions shrink and insertions widen the focus. Other than that, the basic mechanism remains and thus the analogy holds. Since aggregation requires only visibility, the analogical model creates no contradictions, if parts of the illuminated area are illuminated by different spots (say, different wave-length) such that each spot serves to identify the input to its particular cell holding some distinct aggregation function.

**Recursive images**

As there are different implementations, we cannot give a single consistent answer for recursion. However, the projection-screen model can cope with both situations mentioned. The single evaluation step identifies a problem, shuts off projection and replaces it with a still-picture. The pseudo-recursion implemented as limited number of iterations places an additional mirror in a slightly angled position such that each iteration can see the non-recursive portion as well as the result of the last iteration.

---

[4] ) In a variation of the queue in a staircase, one might think of situations where only every n-th step holds an incrementing formula. In this case, the argument raised for deletions obviously applies only if a formula bearing cell is deleted. Otherwise, the many-handed figure case applies.

## 6. Summary

Cell based specification and immediate feedback made spreadsheets a programming device for non-programmers. Spreadsheets provide abstraction through information hiding and modularity. Operations such as copy/paste, drag, and fill support a "next cell"-development approach. While this is convenient when developing a spreadsheet, it is harmful if changes and modifications have to be made. Here, a solid conceptual model is needed.

Current spreadsheet implementations do not strictly follow any of the established conceptual models. They rather follow a teleological approach of "what the user probably intends to do". But phrases containing the word "probably" are problematic as they do not hold for all situations. This poses a challenge for education. If limits and "critical factors" remain unnoticed or misconceived, spreadsheet quality is seriously impacted.

This paper presented a model to explain spreadsheet mechanics to beginners that extends to expert level concepts. It should not be one of the dangerous crutches that break when users try to leave their cradle. Whether this is true has yet to be tested in formal experiments and by exposing the model to the community. After all, with spreadsheet education one has to consider that spreadsheet programmers are not interested in programming per se [Peyton-Jones, Blackwell, Burnett, 2003]. But nevertheless, progress needs education. This holds for spreadsheets as for any other intellectual activity.